\documentclass[aps,prb,showpacs,twocolumn,groupedaddress]{revtex4}

\usepackage{graphicx,bbm}

\def\tr{\,{\rm tr}\,}
\def\ket#1{|#1\rangle}
\def\bra#1{\langle#1|}
\def\braket#1#2{\langle #1 | #2 \rangle}

\begin{document}

\title{Many body localization in Heisenberg XXZ magnet in a random field}

\author{Marko \v Znidari\v c$^1$, Toma\v z Prosen$^1$ and Peter Prelov\v sek$^{1,2}$}

\affiliation{
${}^1$ Department of Physics, FMF, University of Ljubljana, Jadranska 19, SI-1000 Ljubljana, Slovenia\\
${}^2$ Jo\v zef Stefan Institute, Jamova 39, SI-1000 Ljubljana, Slovenia}

\date{\today}

\begin{abstract}
We numerically investigate Heisenberg XXZ spin-$1/2$ chain in a spatially random static magnetic field. 
We find that tDMRG simulations of time evolution can be performed efficiently,
namely the dimension of matrices needed to efficiently represent the time-evolution increases 
linearly with time and entanglement entropies for typical chain bipartitions increase logarithmically. 
As a result, we show that for large enough random fields infinite temperature spin-spin correlation function displays exponential localization in space indicating insulating behavior of the model.
\end{abstract}


\maketitle

\section{Introduction}

Interesting open problems in solid state theory are frequently connected to strongly correlated systems. While the physics of noninteracting particles is well understood much less in known about many body interacting systems. Analytical solutions are typically not possible and numerical calculations are notoriously difficult precisely due to strong correlations. In the theory of quantum information quantum correlations, or entanglement, are one of the central objects of study. Recently,
ideas of quantum information inspired a new numerical method, called time-dependent density-matrix renormalization group (tDMRG)~\cite{MPS,MPO}. The method, originally proposed for simulation of time evolution of quantum systems on classical computers, has been shown rigorously to be efficient in the number of particles at a fixed evolution time~\cite{Osborne:06}. But the more relevant question is what is its time efficiency, {\em i.e.} how the complexity grows with simulation time? Numerical experiments showed that the method is efficient only in some rather special cases~\cite{errors}, while in general it fails~\cite{PRE:07} due to fast growth of entanglement with time. An open question is whether
tDMRG can nevertheless be used efficiently for some nontrivial interacting many particle system and for generic initial conditions? Our aim is to answer this question in affirmative. We show that tDMRG is time efficient for an interacting 1d Heisenberg chain in a disordered (spatially random) magnetic field. This efficiency allows us to calculate infinite temperature correlation functions for large chains of the order of a hundred sites and showing that the many-body states in disordered
Heisenberg model are localized, at least at large enough random
fields.
 
While a single particle Anderson localization is well understood, for a review see~\cite{Kramer:93},
 the interplay between disorder and (strong) interactions in the onset of
localization, as manifested e.g. in vanishing d.c. transport
coefficients, is a subject of an ongoing research, see {\em e.~g.} refs.~\cite{general_mbloc} and references therein. 
The simplest interacting situation is that of two particles. It has been studied for the first time in~\cite{Dima:94} and shown that the localization length can drastically increase in the presence of interaction. This has been confirmed by many subsequent studies. Situation for many interacting particles, {\em e.g.} for finite densities, is much less clear. It has been explored in~\cite{Halfpap:00} by calculating time evolution of wavepackets, showing that the center of mass extension of the wavepacket grows logarithmically with time. Later, the influence of the disorder on the entanglement has been studied for single particle states~\cite{Bambi:04} and for quantum computer simulating single particle localization~\cite{Montangero:04}. Disordered Heisenberg model and entanglement properties of its eigenstates has been studied in~\cite{Lea:04} where a transition in level spacing distribution from Poissonian for no disorder, to Wigner-Dyson distribution of random matrix theory, and back to Poissonian in the case of localization, has been observed as the disorder amplitude is increased. Spectral statistics for interacting disordered system has been also studied in~\cite{Huse:06}, sugesting the existence of localization for sufficiently strong disorder. Characterization of metal-insulator transition in disordered systems
(in absence of interaction) in terms of spectral statistics  have been studied in numerous works, see, {\em e.~g.} ref.~\cite{LSDdis} and references therein. Similar transitions in spectral statistics have been observed also in interacting systems, see e.g. refs.~\cite{LSDint}. In~\cite{Viola:07} a relation between generalized entanglement and inverse participation ratio has been obtained for eigenstates of a disordered Heisenberg model whereas in~\cite{Olivier:07} Meyer-Wallach entanglement has been calculated for random states localized on $M$ random or adjacent computational states. Localization in many body system can be obtained also by constructing special on-site disorder~\cite{Dykman:04}.

\section{Numerical method}

We will use tDMRG method to calculate time evolution of pure states~\cite{MPS}, $\ket{\psi(t)}=U(t) \ket{\psi(0)}$, or time evolution of operators~\cite{MPO}, $O(t)=U^\dagger(t)OU(t)$. Matrix product states (MPS) are used to represent pure states, $\ket{\psi}=\sum_{s_j}\tr{(A_1^{s_1}\cdots A_n^{s_n})}\ket{s_1\ldots s_n}$, where $\ket{s_1\ldots s_n}$ are computational basis states with each $s_j$ taking two values $s_j \in \{0,1\}$, {\em i.e.} local dimension is $d=2$.
%, as well as operators. 
For operators a matrix product operator (MPO) is expansion 
\begin{equation}
O=\sum_{s_j}{\tr{(A_1^{s_1}\cdots A_n^{s_n})} \sigma_1^{s_1}\otimes \cdots \otimes \sigma_n^{s_n}},
\label{eq:O}
\end{equation} 
where a basis of Pauli matrices is used for each site, that is each $s_j$ can now take four different values, $s_j \in \{0,{\rm x},{\rm y},{\rm z}\}$ (local dimension is $d=4$). The advantage of MPS/MPO representation is that the transformation acting on the neighboring spins can be done locally, that is by transforming only two adjacent matrices. Short time propagator $U(\tau)=\exp{(-{\rm i} H \tau)}$ generated by nearest neighbor Hamiltonian $H$ is decomposed using a 2nd order Trotter-Suzuki formula into a series of one and two qubit (qudit) operations. After each application of a two qubit gate the dimension $D$ of the two matrices involved increases by a factor of $d$. In order to prevent the growth of matrix size with time one truncates their size using a singular values decomposition, keeping only the largest singular values. Truncation after application of a single two qubit gate $U_k$ introduces truncation error equal to the sum of squares of the discarded singular values, 
$\eta(U_k)=1-\sum_{j=1}^D\mu^2_j(U_k)$, if $\mu_j(U_k)$ are decreasingly ordered singular values ({\em i.e.} Schmidt coefficients) of the bipartition with the cut being on the bond affected by the gate $U_k$. Total truncation error after application of a series of gates, $U(t)=\prod_{k} U_k$, is then the sum of individual errors
$\eta_{\rm tot}(t)=\sum_k \eta(U_k)$.
Truncation error $\eta_{\rm tot}(t)$ scales with the time step $\tau$ as $\eta_{\rm tot}(t) \propto \tau$. By using Trotter-Suzuki factorization of
$U(t)$ we also introduce Trotter error. For our choice of 2nd order formula the error in fidelity $1-|\braket{\psi_{\rm tDMRG}}{\psi_{\rm exact}}|^2$ scales as $\propto \tau^6 (t/\tau)^2=\tau^4 t^2$ (note that 
the ``phase'' error $|\braket{\psi_{\rm tDMRG}}{\psi_{\rm exact}}-1|$ scales as $\propto \tau^2 t$). If one starts evolution from a product state/operator at $t=0$, which is always the case in our simulations, the error is initially dominated by the Trotter error but for larger times the precision of tDMRG is eventually determined by the truncation error $\eta_{\rm tot}$. In the following we are going to focus solely on the truncation error.   

Throughout this paper, when calculating evolution of pure states, we start with a random 
product state, {\em i.e.} $\ket{\psi(0)}=\ket{\psi_1}\otimes \cdots \otimes\ket{\psi_n}$, where 
$\ket{\psi_j}$ is a random state of $j$-th qubit corresponding to a random point on its Bloch sphere. Note that in numerical simulations, in order to obtain an infinite temperature behaviour, we average over initial random states.
When simulating Heisenberg evolution of operators - which will be used for computation of spin-spin correlations - the initial operator will be chosen to be the spin projection at a quarter of the lattice
 $O(0) =  s^{\rm z}_{n/4} $. The initial state is therefore in both cases, of operator or pure state dynamics, separable. However, time evolution is expected to produce entanglement. 
In all numerical computations, we also average over random realizations of disorder.

The system we study is a 1d spin-$1/2$ Heisenberg XXZ model in a random magnetic field,
\begin{equation}
H=\sum_{j=1}^{n-1}\left(s_j^{\rm x} s_{j+1}^{\rm x}+s_j^{\rm y} s_{j+1}^{\rm y}+\Delta s_j^{\rm z} s_{j+1}^{\rm z}\right)+\sum_{j=1}^n h_j s_j^{\rm z}.
\label{eq:H}
\end{equation}
where $s^\alpha_j$ are canonical spin $1/2$ variables.
The magnetic field will be chosen randomly and uniformly in the interval $[-h,h]$. 
%We shall take $h=5$ and $\Delta=0.5$ for most of
%numerical results shown in this paper, a
%throughout the paper, apart from Fig.~\ref{fig:freeF} where we use $\Delta=0$ and $h=1$. 
The case of $\Delta=0$ is a special case and should be clearly distinguished from nonzero $\Delta$. Using Wigner-Jordan transformation, Hamiltonian (\ref{eq:H}) with $\Delta=0$ can be transformed to a bilinear fermionic system 
$\sum_j (c_{j+1}^\dagger c_j + h.c. +h_j n_i)$ with $n_i=c_i^\dagger
c_i$ which represents the model of noninteracting spinless fermions
with diagonal disorder known to exhibit Anderson localization in
1d. The case $\Delta \neq 0$ introduces the interaction or
correlations among electrons (through $\Delta n_j n_{j+1}$) which can
qualitatively change properties of the system.
The aim of this letter is twofold: (i) to study localization in an interacting disordered system, and (ii) to show that time evolution with tDMRG, as opposed to non-integrable and non-disordered situation, is efficient for such a system.

The computational complexity of simulating quantum evolution on a classical computer using tDMRG is determined by the growth of bipartite entanglement. If bipartite entanglement increases with time we have to increase dimension of matrices $D$ with time in order to prevent truncation error $\eta_{\rm tot}(t)$ from growing. Because the number of computer operations in tDMRG scales as $\sim D^3$ it is crucial  to know how fast do we have to increase $D$? We are going to study necessary dimension $D_\varepsilon(t)$ in order for the truncation error at time $t$ to be less than $\varepsilon$. If the necessary $D_\varepsilon(t)$ grows linearly with time we say that the simulation is efficient and if it grows exponentially simulation is inefficient. For a {\em quantum chaotic} many body system $D_\varepsilon(t)$ grows exponentially with time~\cite{PRE:07}. Before looking at $D_\varepsilon(t)$ let us have a look at bipartite entanglement. Note that the z-component of the total spin
$S^{\rm z}=\sum_j s^{\rm z}_j$ is a conserved quantity $[H,S^{\rm z}]=0$. In the following we
will consider $T\to \infty$ limit of the model, so that random states
will on the average represent states with zero polarization, $S^{\rm z}=0$.
In the fermion representation this corresponds to the half-filled band
with $n/2$ spinless particles where the effect of interactions is
strongest. In the other extreme case $S^{\rm z}=n/2-1$ the
interaction $\Delta$ term is constant and therefore the problem is
equivalent to 1d Anderson model of localization without interaction.
 
\section{Results}

\subsection{Entanglement entropy}

\begin{figure}[h!]
\includegraphics[width=0.65\linewidth,angle=-90]{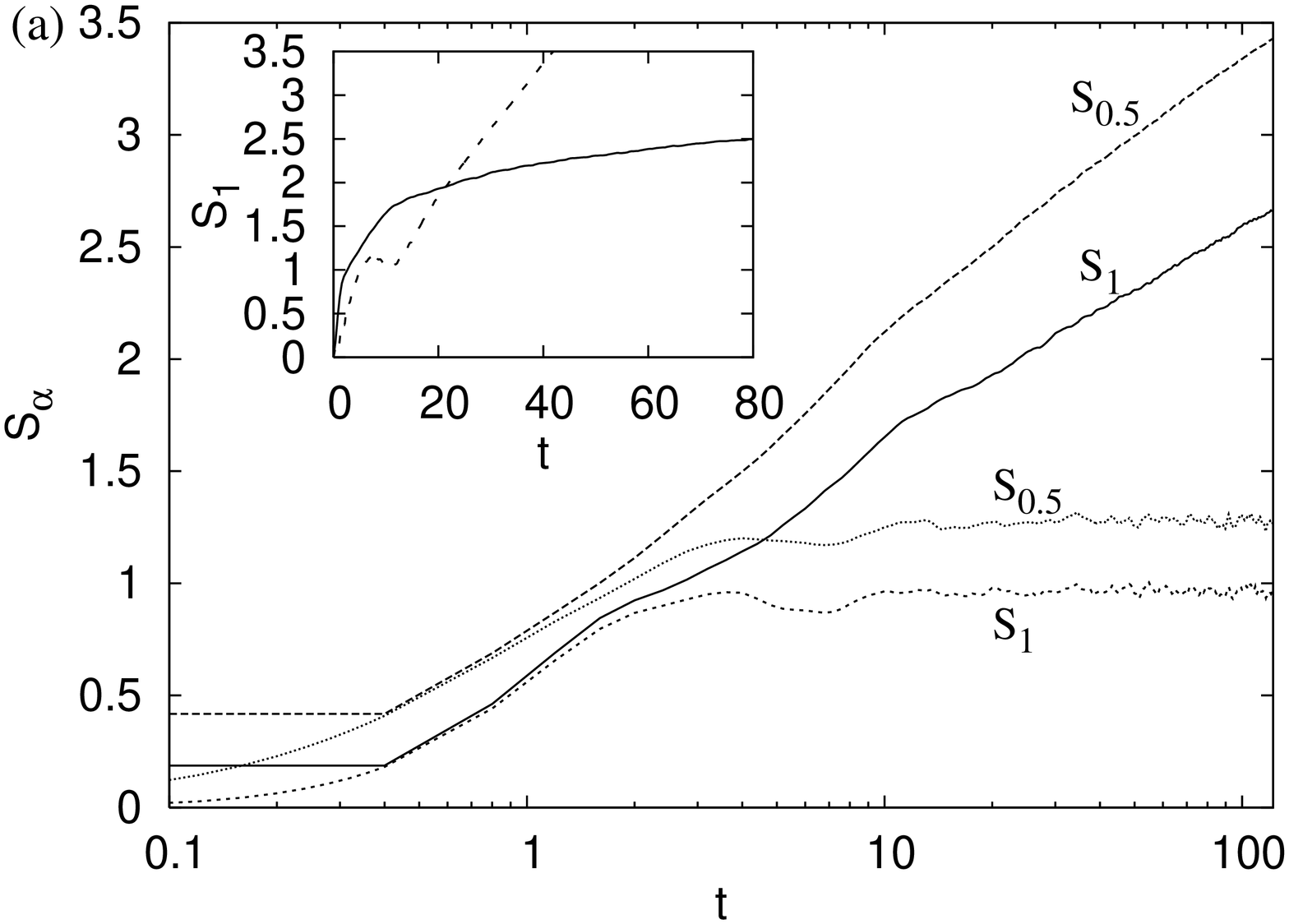}
\hskip -10mm
\includegraphics[width=0.65\linewidth,angle=-90]{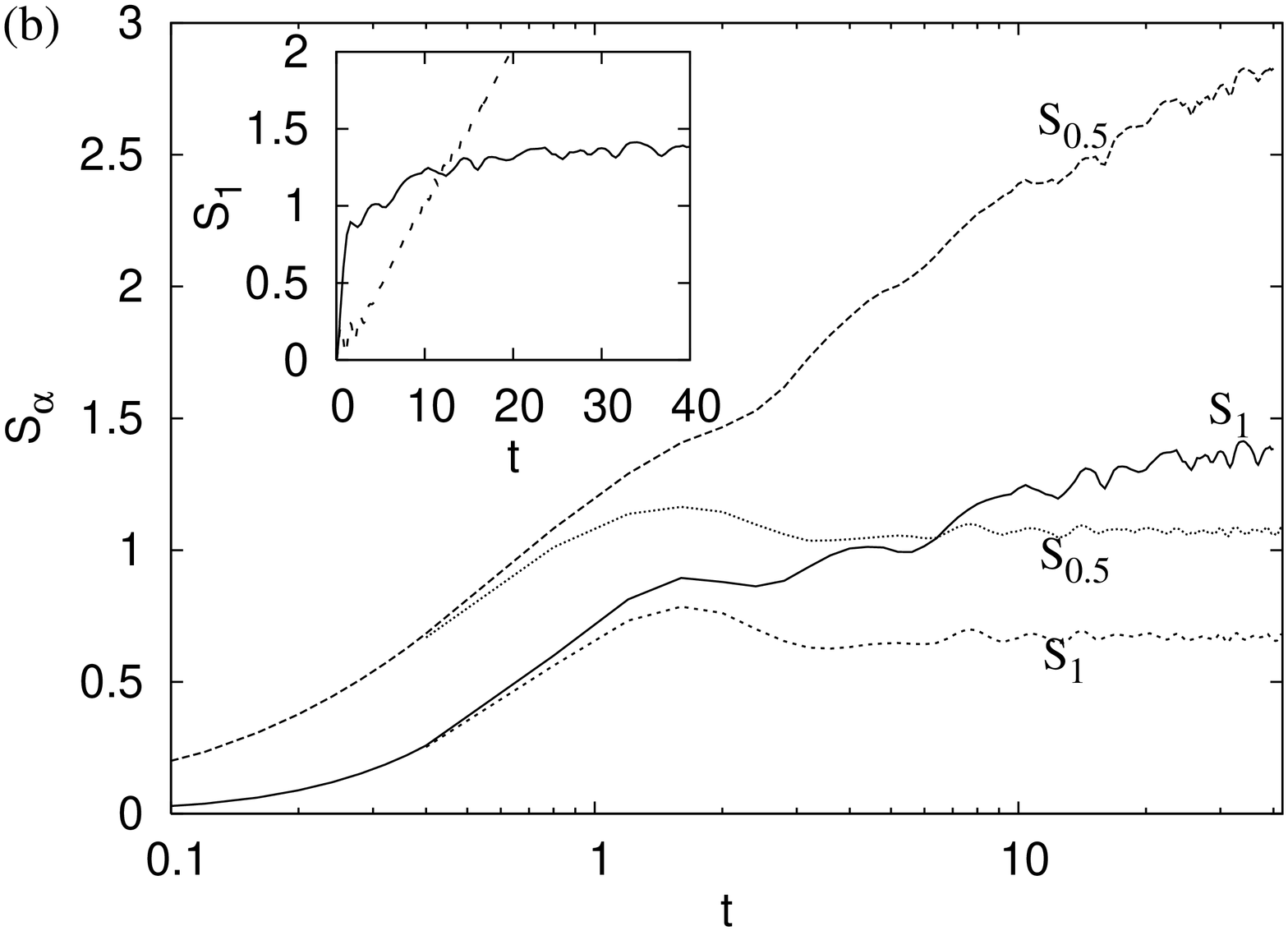}
\hfill
\caption{Entanglement entropies $S_{0.5}(t)$ and $S_1(t)$ for pure state evolution (a), and operator
space evolution (b) [see text for details]. We compare cases of $\Delta=0.5$ (growing curves) and $\Delta=0$ (saturating curves),  for $n=50$, average over 100 (a), 21 (b, for $\Delta=0.5$), 1000 (b, for $\Delta=0$) disorder 
realizations with magnetic field magnitude $h=5$.
In insets we compare {\em logarithmic} growth of $S_1(t)$ in disordered case (full curves, same data as
in main frames)  to {\em linear} growth in the case of staggered magnetic field $h_j = (-1)^j 5/\sqrt{3}$
and $\Delta=0.5$ (dashed curves).
}
\label{fig:Pn_S} 
\end{figure}
The amount of bipartite entanglement can be measured by von Neumann entropy 
$S_1(t)=\tr{(\rho_A \log_2{\rho_A})}$, or in general by the Renyi entropy, of the reduced density matrix
\begin{equation}
S_\alpha(t)=\frac{\log_2{\tr{\rho_A^\alpha}}}{1-\alpha},\qquad \rho_A=\tr_{B}{\ket{\psi(t)}\bra{\psi(t)}}.
\label{eq:Reny}
\end{equation}
We shall always consider bipartite cut with the first $m$, or last $n-m$, qubits constituing subspace $A$, or $B$, respectively. In the case of operators, expansion coefficients in the basis of Pauli matrices (\ref{eq:O}) 
are treated as expansion coefficients of a {\em super-ket} in a Hilbert space of dimension $4^n$, for which the {\em operator-space entanglement entropy} is then calculated. For the discussion of approximability of states with MPS form and its relation to Renyi entropies see~\cite{Schuch:07}. We show in Fig.~\ref{fig:Pn_S} time dependence of Renyi $S_{0.5}(t)$ and von Neumann $S_1(t)$ entropies. For simulation of pure states, Fig.~\ref{fig:Pn_S}a,
 the entropies can be seen to grow {\em logarithmically} with time in the interacting case $\Delta=0.5$, whereas they {\em saturate} to a constant for $\Delta=0$ which seems consistent with an Anderson quasi-particle localization. Entropies in
 Fig.~\ref{fig:Pn_S} are shown for the `worst case' bipartition only,  i.e. for $m$ which maximizes them, whereas results are qualitatively equivalent for half-half bipartition ($m=n/2$), or average bi-partition (average over $m=1\ldots n-1$).
For evolution of operators, Fig.~\ref{fig:Pn_S}b, in the interacting case $\Delta=0.5$, $S_{0.5}(t)$ again grows logarithmically with time whereas the growth of $S_1(t)$ is slower than logarithmic, perhaps saturating. In fact, saturation of operator space entropies $S_\alpha(t)$, for large enough $\alpha>\alpha^*$, seems a probable interpretation of our results since it is compatible with localization of
correlation functions reported below (see Fig.~\ref{fig:intFs}).
Slower growth of entanglement entropies (or even saturation) for operators as compared to faster growth for pure states is consistent with a qualitatively similar finding for a homogeneous transverse Ising model~\cite{PRE:07}. In insets of Fig.~\ref{fig:Pn_S} we compare our data to tDMRG simulation
for a non-disordered model (\ref{eq:H}) in a staggered field of comparable strength 
$h_j=(-1)^j h/\sqrt{3}$, and there we find a {\em linear} growth of $S_\alpha(t)\propto t$, for pure states and operators, which is consistent with an {\em exponential} increase of $D_\epsilon(t)$ found for 
non-integrable models \cite{PRE:07}. Note that there is in general no simple relation between the behavior of pure state entanglement entropy and operator-space entanglement entropy.
Only in the special case of rank-one projection operators $O=\ket{\psi}\bra{\psi}$ we find a simple relation,
namely that operator-space von Neumann entanglement entropy of $\ket{\psi}\bra{\psi}$ equals two times entanglement entropy of $\ket{\psi}$. Hence the operator-space entanglement entropy is also {\em not equivalent}
to an entanglement of a mixed state.

\begin{figure}[h!]
\includegraphics[width=0.65\linewidth,angle=-90]{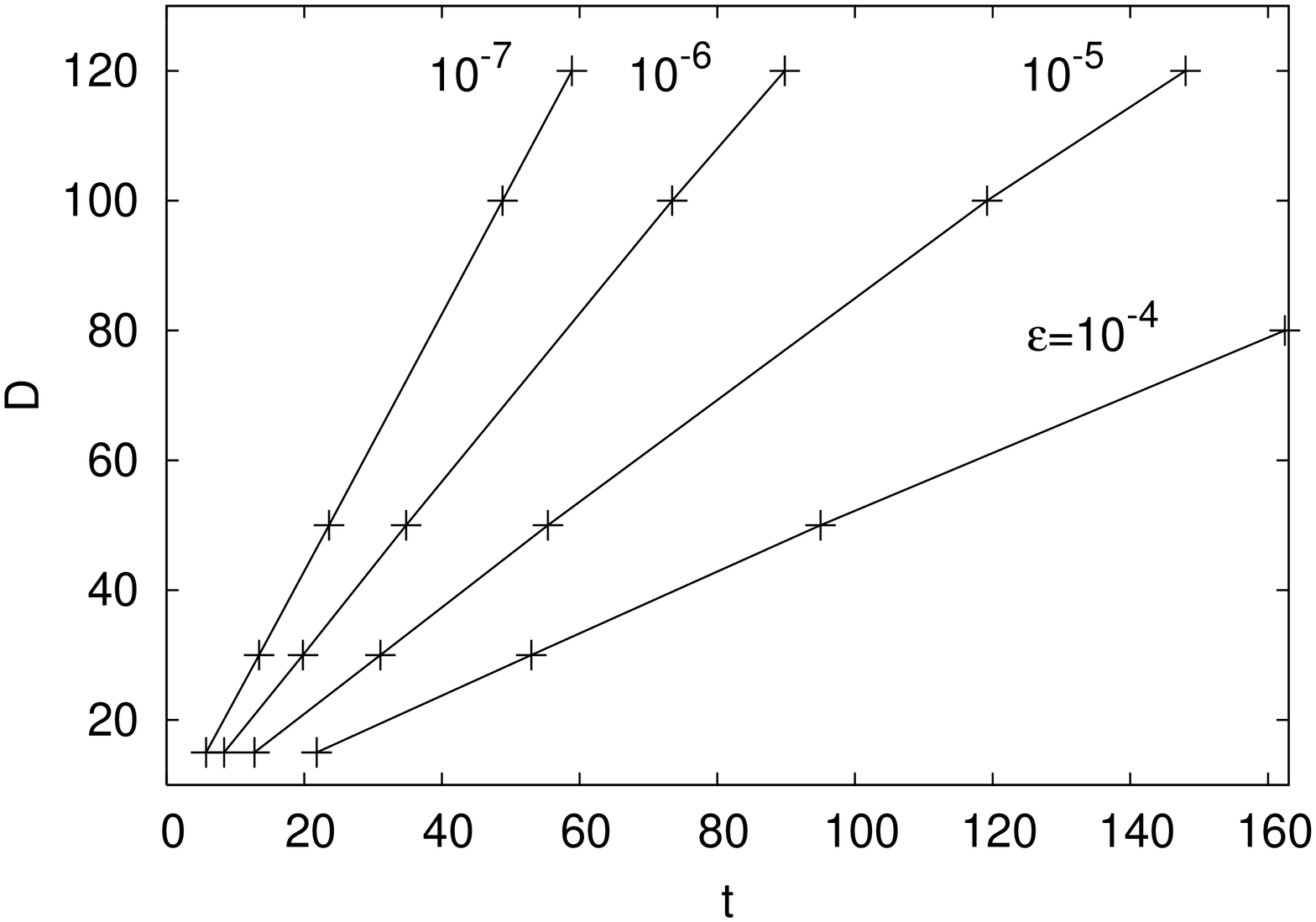}
\hskip -10mm
\includegraphics[width=0.65\linewidth,angle=-90]{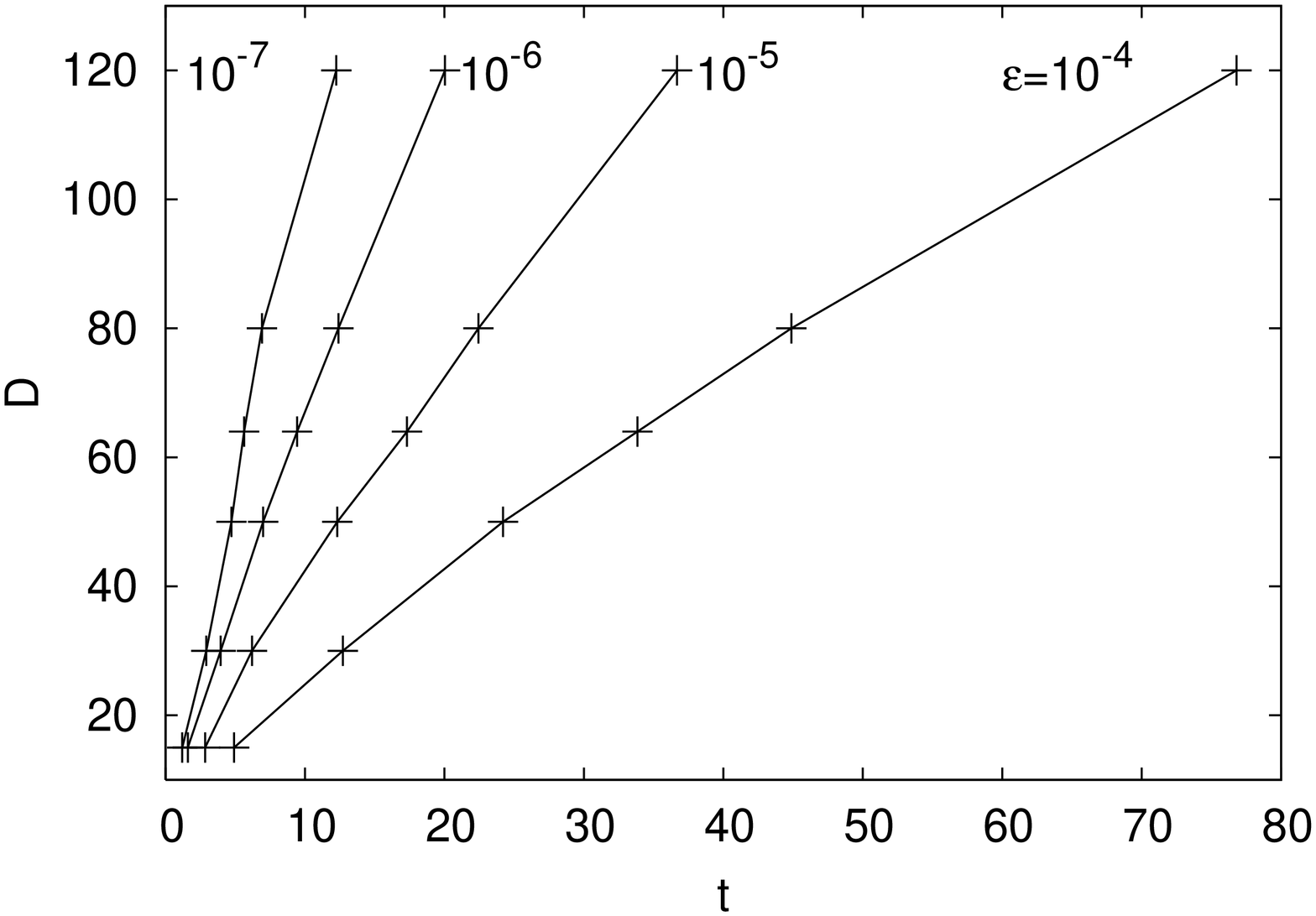}
\caption{Dimension $D_\varepsilon(t)$ for pure state (top) and operator evolution (bottom). Data for various $\varepsilon$ clearly indicate linear growth with time. $\Delta=0.5$, $n=50$, and a single disorder realization with $h=5$ (the same realization for all data points).}
\label{fig:Pn_Dodt} 
\end{figure}

As we have seen, entropies grow at most logarithmically for a disordered field. This gives us hope that the evolution with 
tDMRG is in fact efficient, meaning that $D_\varepsilon(t)$ grows polynomially with time. This is indeed the case as shown in Fig.~\ref{fig:Pn_Dodt}. The necessary dimension of matrices grows linearly with time therefore the simulation of disordered Heisenberg chain is efficient, both for pure states and for operators. This must be contrasted to the other known efficients case of 
tDMRG, namely integrable transverse Ising model~\cite{PRE:07}, where the simulation is efficient only for operators.

\begin{figure}[h!]
\includegraphics[width=0.65\linewidth,angle=-90]{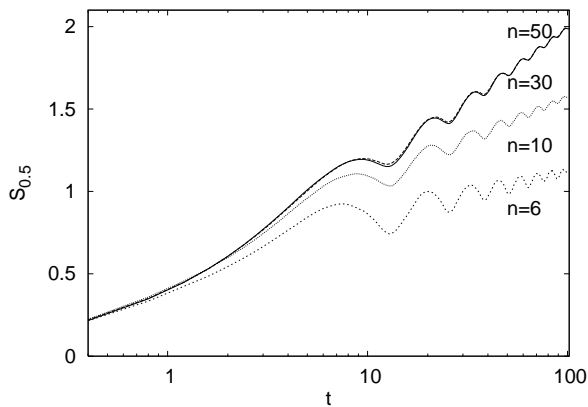}\hfill
\caption{Renyi entanglement entropy growth $S_{0.5}(t)$ for pure state evolution and different system sizes $n$. We show numerical results, averaged over $100$ realizations of disorder, for $n=6$, $10$, $30$ and $50$, with short-dashed, dotted, long-dashed, and full curves, respectively. Note that $n=30$ and $n=50$ cases almost overlap. In all cases $\Delta=0.5$ and $h=5$ and we show the average entropy over all bipartite cuts in order to have smoother data.}
\label{fig:Pnn_S} 
\end{figure}

We note that in the numerical results presented above there is no significant finite size effects. In order to demonstrate
that, we plot in Fig.~\ref{fig:Pnn_S} the growth of Renyi entropy $S_{0.5}(t)$ for different system sizes
$n$, and observe significant finite size effects only for size smaller than $30$. 

\subsection{Infinite temperature correlation function}

After establishing that tDMRG simulation of a disordered Heisenberg chain is efficient we want to calculate some physically relevant quantity. We choose an infinite temperature 
spin-spin correlation function $C(r,t)$,
\begin{equation}
C(r,t)=2^{-n}\tr\left\{s^{\rm z}_{n/4}(t) s^{\rm z}_{n/4+r}\right\},
\label{eq:C}
\end{equation}
which is computed from MPO representation of Heisenberg dynamics $s^{\rm z}_{n/4}(t)$.
By calculating $C(r,t)$ we can directly 
assess the many-body localization. In particular, $C(r,t\to
\infty)$ gives direct information on d.c. transport properties, i.e.,
whether the d.c. spin diffusion constant is finite, or equivalently
for fermions, whether the system is a conductor (normal resistor) or
Anderson insulator (at finite $T$). 
First, we are going to consider noninteracting system $\Delta=0$. The reason to study this case is to verify recently obtained bounds on the propagation of correlations in disordered systems~\cite{Osborne:07}. It has been proven that the correlations are exponentially suppressed outside an effective logarithmic lightcone.  
\begin{figure}[h!]
\includegraphics[width=0.6\linewidth,angle=-90]{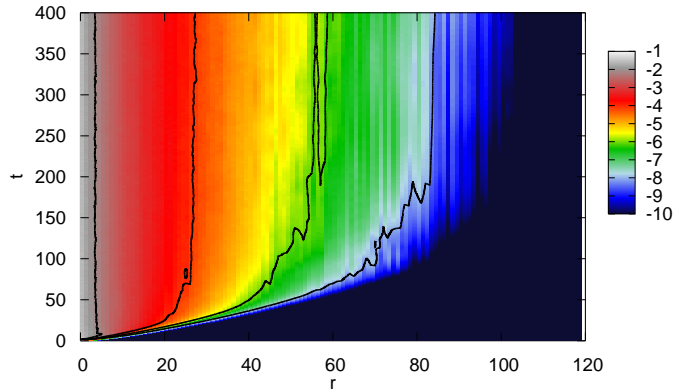}\hfill
\caption{(Color online) Spin-spin correlation function $C(r,t)$ ($\log_{10}$ color code) for non-interacting case, $\Delta=0$, $n=500$, average over 1000 disorder realizations with $h=1$. Solid curves are isolines at $10^{-8}$, $10^{-6}$, $10^{-4}$ and $10^{-2}$, from right to left.}
\label{fig:freeF} 
\end{figure}
Note that since this case $\Delta=0$ is equivalent to noninteracting fermions the necessary dimension of matrices $D$ actually saturates~\cite{PRE:07} at $D_\varepsilon(t)=4$, meaning that the calculation is very efficient and chains of thousands of spins are easily accessible. From the calculated correlation function shown in Fig.~\ref{fig:freeF} we can see that the correlation function decays exponentially in space, and is frozen in time after sufficiently long time, $|C(r,t)| < K \exp(-r/\xi)$, for some $K,\xi > 0$.
This means that the rigorous estimate \cite{Osborne:07} is in fact over-pessimistic and
can perhaps be improved~\cite{private}.
We actually observe three regimes: (i) for small times correlations propagate balistically, seen as a linear growth of isocurves $C(r(t),t) = {\rm const}$. (ii) After that we have a logarithmic propagation of correlations, reflected in a log-shape of isocurves. (iii) For large time, localization sets in and the spatial correlation function gets its asymptotic exponential shape. The crossover times between the regimes increase with decreasing value of correlation on the isocurve. 

\begin{figure}[h!]
\includegraphics[width=0.65\linewidth,angle=-90]{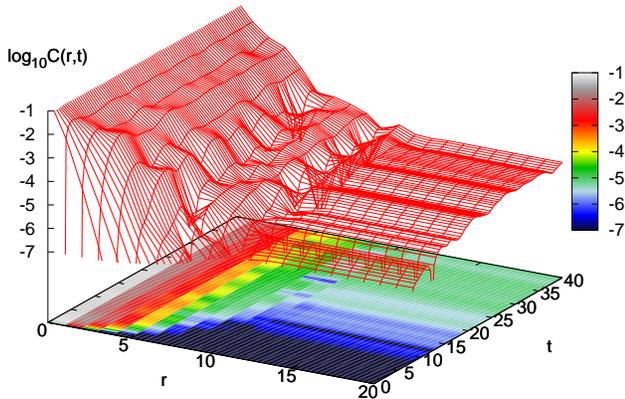}\hfill
\caption{(Color online) Correlation function $C(r,t)$ for $n=50$, $\Delta=0.5$, averaged over 21 disorder realization with $h=5$.}
\label{fig:intFs} 
\end{figure}

In Fig.~\ref{fig:intFs} correlation function is shown for the interacting system, $\Delta=0.5$. Interestingly, we again observe localization with qualitatively similar structure of correlation function $C(r,t)$ as in the case $\Delta=0$. Note that numerical simulation is in this case much harder. For the case shown the dimension of matrices has been $D=64$ which, as can be inferred from Fig.~\ref{fig:Pn_Dodt}, means truncation error $\eta_{\rm tot}(t \approx 40)\approx 10^{-3}$. Because of truncation errors, for large times $t>20$ and distances $r>10$ one can see sort of a plateau in the correlation function in Fig.~\ref{fig:intFs} which can be systematically decreased by increasing $D$. 
Our results indicate that the interaction does not destroy localization, at least for large disorder.
This appears to be in contrast with the predicted metal-insulator transition at a finite $T$ in the corresponding fermionic system~\cite{Altshuler}. 

However, in light of observed logarithmic growth of entanglement entropies with time we cannot
exclude another possibility of very small diffusion constant and, in strict sense, absence of localization.
Yet, another possibility would be existence of localization delocalization transition at smaller disorder
strength $h$, but this regime is much harder to simulate with the present method as entanglement entropies increase faster for smaller $h$.

Still, we were able to verify that the entropy grows logarithmically also for smaller fields, {\em e.g.} for $h=2$, as well as for other interaction strength, {\em e.~g.} $\Delta=1.5$. For example, for $\Delta=0.5$ and $h=2$ we find $S(t) \sim c \log t$ with $c$ which is about $4$ times larger than for $h=5,\Delta=0.5$. This indicates that the system probably displays the same kind of localization also for smaller magnetic field strength. If this is the case, transition in level spacing statistics observed in~\cite{Lea:04,Huse:06} might be just due to finite size effects (localization length being larger than the system size).   

We have also checked that the results of correlation functions, in the regimes which we show, do not significantly depend on the system size $n$. For instance, the correlation function for $n=20$ is practically the same as the one for $n=50$ in Fig.~\ref{fig:intFs}. 

\section{Conclusions} 

tDMRG is for general systems inefficient due to fast entanglement growth. Nevertheless, we have shown that it can be efficient for a disordered Heisenberg model, and potentially as well for some other disordered interacting models in one dimension. Spin-spin correlations evaluated in our study at large
temperatures exhibit localization in spite of nontrivial interaction
term, inferring that all-many body states are localized in large enough disorder.  

It should be noted that this is the first successful application of tDMRG for simulation of long time evolution at high
temperature in a non-solvable (non-integrable) system. The efficiency of the method is related to an interesting observed logarithmic asymptotic growth of entanglement entropy for time evolution of, both, pure states and operators.
 
We acknowledge support by Slovenian Research Agency, program P1-0044, and grant J1-7437.


\begin{thebibliography}{1}

\bibitem{MPS} G.~Vidal, Phys.~Rev.~Lett. {\bf 91}, 147902 (2003); {\em ibid.} {\bf 93}, 040502 (2004); S.~R.~White and A.~E.~Feiguin, Phys.~Rev.~Lett.{\bf 93}, 076401 (2004); A.~J.~Daley {\em et al.}, J.~Stat.~Mech. {\bf 4}, P04005 (2005).

\bibitem{MPO} F.~Verstraete {\em et al}, Phys.~Rev.~Lett. {\bf 93}, 207204 (2004); M.~Zwolak and G.~Vidal, Phys.~Rev.~Lett. {\bf 93}, 207205 (2004); A.~E.~Feiguin and S.~R.~White, Phys.~Rev.~B {\bf 72}, 220401(R) (2005).

\bibitem{Osborne:06} T.~J.~Osborne, Phys.~Rev.~Lett {\bf 97}, 157202 (2006).

\bibitem{errors}
D.~Gobert {\em et al.}, Phys.~Rev.~E {\bf 71}, 036102 (2005).
 

\bibitem{PRE:07} T.~Prosen and M.~\v Znidari\v c, Phys.~Rev.~E {\bf 75}, 015202(R) (2007).

\bibitem{Kramer:93} B.~Kramer and A.~Mac{K}innon, Rep.~Prog.~Phys. {\bf 56}, 1469 (1993).

\bibitem{general_mbloc}
C.~A.~Doty and D.~S.~Fisher, Phys. Rev. B {\bf 45}, 2167 (1992);
P.~Schmitteckert {\em et al.}, Phys. Rev. Lett. {\bf 80}, 560 (1998);
D.~Weinmann {\em et al.}, Eur.~Phys.~J. B {\bf 19}, 139 (2001);
C.~Schuster {\em et al.}, Phys.~Rev.~B {\bf 65}, 115114 (2002); 
L.~Urba and A.~Rosengren, Phys. Rev. B {\bf 67}, 104406 (2003);
J.~M.~Carter and A.~MacKinnon, Phys. Rev. B {\bf 72}, 024208 (2005);
G.~De Chiara {\em et al.} J. ~Stat.~Mech.{\bf 5}, P03001 (2006).

\bibitem{Dima:94} D.~L.~Shepelyansky, Phys.~Rev.~Lett. {\bf 73}, 2607 (1994).

\bibitem{Halfpap:00} O.~Halfpap, PhD thesis, Universit\" at Hamburg (2000); O.~Halfpap {\em et al.},  Ann.~Phys-Berlin {\bf 7}, 483 (1998); X.~Waintal {\em et al.}, Eur.~Phys.~J. B {\bf 7}, 451 (1999).

\bibitem{Bambi:04} H.~Li, X.~Wang and B.~Hu, J.~Phys.~A {\bf 37}, 10665 (2004).

\bibitem{Montangero:04} S.~Montangero, Phys.~Rev.~A {\bf 70}, 032311 (2004).

\bibitem{Lea:04} L.~F.~Santos {\em et al} Phys.~Rev.~A {\bf 69}, 042304 (2004); L.~F.~Santos, J.~Phys.~A {\bf 37}, 4723 (2004).

\bibitem{Huse:06} V.~Oganesyan and D.~A.~Huse, Phys.~Rev.~B {\bf 75}, 155111 (2007).

\bibitem{LSDdis} B.~I.~Shklovskii {\em et al}, Phys.~Rev.~B {\bf 47}, 11487 (1993); A.~D.~Mirlin, Phys.~Rep. {\bf 326}, 259 (2000).

\bibitem{LSDint} G.~Montambaux {\em et al}, Phys.~Rev.~Lett. {\bf 70}, 497 (1993); R.~Berkovits, Europhys.~Lett. {\bf 25}, 681 (1994).

\bibitem{Viola:07} L.~Viola and W.~G.~Brown, {\tt e-print} quant-ph/0702014.

\bibitem{Olivier:07} O.~Giraud {\em et al}, {\tt e-print} arXiv/0704.2765.

\bibitem{Dykman:04} M.~I.~Dykman {\em et al.}, {\tt e-print} cond-mat/0401201.

\bibitem{Schuch:07} N.~Schuch {\em et al.}, {\tt e-print} arXiv/0705.0292.

\bibitem{Osborne:07} C.~K.~Burrell and T.~J.~Osborne, Phys.~Rev.~Lett. {\bf 99}, 167201 (2007).

\bibitem{private} Indeed, the general bound in~\cite{Osborne:07} can be improved for the particular correlation function studied here. C.~K.~Burrell and T.~J.~Osborne, {\em private communication}. 

\bibitem{Altshuler} D.~M.~Basko {\em et al.}, {\tt e-print} cont-mat/0602510; Annals of Physics {\bf 321}, 1126 (2006). 

\end{thebibliography}
\end{document}